# Mixtures of Diethyl Sulfoxide and Methanol: Structure and Thermodynamics

Vitaly V. Chaban[1] and Nadezhda A. Andreeva[1,2]

(1) P.E.S., Vasilievsky Island, Saint Petersburg, Russian Federation.

(2) Department of Physics, Peter the Great Saint Petersburg State Polytechnic University, Russian Federation.

**Abstract**. Mixtures of sulfoxides with molecular solvents possess interesting physical-chemical properties and may have applications in chemical synthesis. Hereby we confirm and rationalize the previously reported excellent miscibility of diethyl sulfoxide (DESO) with methanol (MeOH). By performing a comprehensive potential energy surface investigation we identified a global minimum for each system and a significant number of local minima described at the hybrid density functional level of theory. A strong 0.18-nm-long hydrogen bond forming between the oxygen atom of DESO and the polarized hydrogen atom of MeOH was evidenced both via the structural and spectral analyses. Our results robustly explain negative deviations in the DESO-MeOH mixtures from ideal behavior and interpret the experimental observations with microscopic precision.

**Graphical Abstract**

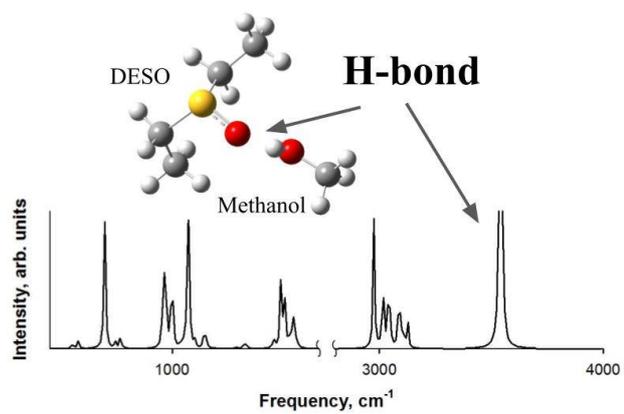

**Introduction**

Being one of the least toxic liquids for animals and human beings, dimethyl sulfoxide (DMSO) has found its wide and well-established applications in solution chemistry, synthesis, medicine, and biological research.[1–10] Senior sulfoxides are presently under investigation.[11–16] Certain evidence exists that diethyl sulfoxide (DESO) performs better in preserving the membrane potential of the living cells as compared to DMSO.[17] It would be reasonable to hypothesize that longer-alkyl-chain sulfoxides demonstrate a higher thermodynamic affinity to the phospholipid bilayer. Therefore, they can more easily penetrate the membranes that can be useful in the context of drug delivery vehicles. Unfortunately, the studies along the mentioned lines are virtually absent. Triolo and coworkers[11] provided an interesting investigation of dibutyl sulfoxide using a computational approach and X-ray diffraction near the room temperature, 320 K, at ambient pressure. Bearing a strong dipolar moiety and medium-length non-polar chains, dibutyl sulfoxide exhibits a certain degree of structural nanometer-scale differentiation that ultimately determines its liquid structure.[11]

Investigation of binary solvents that contain DESO is underway the most popular being DESO-water mixtures.[4,15,16,18,19] It was revealed and convincingly rationalized that DESO is miscible with water at all compositions and exhibits a strongly favorable free energy of hydration. Hydration in this case is fostered by the formation of hydrogen bonds between the hydrogen atom of water and the oxygen atom of DESO. Computer simulations of dialkyl sulfoxides have recently been reported and provided important physical insights at the atomistic level of precision.[11,14]

In the present work, we report potential energy surfaces of the diethyl sulfoxide (DESO)-methanol(MeOH) mixtures and exemplify the most thermodynamically stable configurations that are responsible for the condensed-phase physical-chemical properties of the mentioned systems. The study of physical-chemical properties of binary liquid mixtures should be considered to be of essential importance to shed light on the non-covalent interactions between common molecular solvents. In their comprehensive study of volumetric properties, Markarian and coworkers recently presented mass densities, apparent molar volumes, partial molar volumes, and excess molar volumes for the DESO-MeOH mixtures over the whole range of compositions.[20] Negative deviations from ideality were clearly observed thanks to strong electrostatic interactions between polar moieties of DESO and MeOH. A hydrogen bond formation involving the oxygen atom of the sulfoxide group and the hydrogen atom of the hydroxyl group, O(DESO)...H(MeOH) was hypothesized, but not characterized. It must be noted that DMSO-alcohol mixtures[10,21–25] are a way better investigated as compared to the DESO-alcohol mixtures. Wu and coworkers published comprehensive sets of data of thermal conductivities in ethanol-dimethyl sulfoxide mixtures that are certainly useful to interpret intermolecular interactions in the alcohol-sulfoxide liquid systems[24] and motivated our present simulations, Such liquid systems represent a particular interest to physical chemists since their constituents form homo and hetero associates.[19,25]

**Methodology**

The investigated systems along with their key parameters are provided in Table 1. One hundred and more input molecular configurations were used to obtain geometry optimized configurations to further analyze their potential energies and structure parameters. All these configurations were selected in a systematic way using molecular dynamics trajectory with periodic thermal external perturbations. First, each system was equilibrated at 300 K using the PM7 Hamiltonian.[26–29] The system was considered properly equilibrated when the block-averaged potential energy and total energy stopped to decrease, whereas temperature stopped to drift and became equal to the reference value. The constant temperature of the system was maintained according to the method of Andersen[30] with a collision frequency of 0.05 implemented in the Atomic Simulation Environment.[31] The equations of motion were solved according to the Verlet algorithm an integration time-step being equal to $5\times10^{-4}$ ps. Due to the small sizes of the systems, the equilibration-stage calculations took less than 20 ps. Consequently, a comprehensive sampling (statistical data accumulation) of the phase trajectory was performed.

Table 1. The parameters of the systems used to sample potential energy surfaces, derive radial distribution functions, and compute infrared spectra. The lengths of molecular dynamics trajectories used for sampling are provided.

| # | n (DESO) | n (MeOH) | n (atoms) | n (electrons) | Trajectory, ps |
|---|----------|----------|-----------|---------------|----------------|
| 1 | 1 | 0 | 16 | 58 | spectrum* |
| 2 | 0 | 1 | 6 | 18 | spectrum* |
| 3 | 1 | 1 | 22 | 76 | 50 + spectrum |
| 4 | 2 | 2 | 44 | 152 | 300 |
| 5 | 3 | 3 | 66 | 228 | 300 |
| 6 | 1 | 8 | 64 | 202 | 600 |
| 7 | 4 | 1 | 70 | 250 | 500 |

- No molecular dynamics simulations were conducted. The systems were used for the spectrum calculations.

Periodically (every 0.5 ps in system 3 and every 4 ps in systems 4-7) a portion of additional kinetic energy equivalent to 1000 K was introduced to the system. This external energy (perturbation) was assigned to atoms to obtain the Maxwell-Boltzmann distribution. The perturbed system was simulated during 0.05 ps after which an immediate set of Cartesian coordinates was written down to the hard disc. This protocol serves the goal of obtaining more diversified profiles of the system before geometry optimization according to the steepest-descent algorithm. It is important to deal with generally physical molecular configurations although with a certain degree of random thermal perturbation. In total, at least 100 immediate configurations of atoms were prepared.

The selected configurations were optimized using the PM7 Hamiltonian[26–29] until the convergence criterion of $10^{-1}$ kJ/mol/Ångstrom was satisfied. The ultimate energy minimization (threshold of $10^{-3}$ kJ/mol/Ångstrom for system 3 and threshold of $10^{-2}$ kJ/mol/Ångstrom for systems 4-7) was performed in the framework of hybrid density functional theory. The Becke-3-Lee-Yang-Parr, B3LYP,[32,33] method was used, whereas the molecular wave function of each system was constructed out of the functions available in the Pople-type polarized atom-centered split-valence triple-zeta basis set 6-31G(d). If the configuration contained negative vibrational frequencies, it was disregarded. If the potential energy of the configuration was less than 4 kJ mol$^{-1}$ different from that of the already obtained configuration of the same system, it was disregarded. The dispersion supplement to improve non-covalent interactions (London forces) was applied.[34]

We used radial distribution functions (RDFs) to confirm the structural patterns which were identified via the comprehensive potential energy surface scan. Vice versa, the peaks of the RDF were used as a control to ensure that we obtained all essential local-minimum molecular configurations. RDFs and the global minimum search procedure supplement one another, while the latter can identify less evident configurations. The presented RDFs correspond to the room-temperature conditions, 300 K.

Infrared spectra for DESO, MeOH and the pair of these molecules (systems 1-3, Table 1) were generated based on the vibrational frequencies of the optimized electronic wave functions. The electronic wave functions were obtained at the B3LYP hybrid density functional theory in conjunction with the 6-311++G** basis set. Polarization and diffuse functions were added to all atoms. The molecular wave function convergence criterion was set to $10^{-4}$ kJ mol$^{-1}$. The geometries were optimized employing the steepest-descent technique. The geometry convergence criterion was set to $10^{-3}$ kJ/mol/Ångstrom. After convergence of the geometry optimization algorithm, vibrational frequencies were computed. No negative frequencies in the profile were identified meaning that a local-maximum molecular configuration was attained.

The described numerical calculations were performed in GAMESS-2014 (electronic-structure calculations),[35] MOPAC-2016 (semiempirical calculations),[26–28] PM7-MD (semiempirical molecular dynamics trajectory),[36] and GMSEARCH (global minimum search)[37] computational chemistry research programs. The functions, procedures and interfaces of the ASE and SciPy.org computational chemistry libraries[31,38,39] were used for certain procedures. Visualization of atomic configurations and preparation of molecular graphics were performed in VMD-1.9.1,[40] Gabedit-2.8,[41] and Avogadro-1.2.0 programs.[42]

**Results and Discussion**

The complexity of the potential energy surface is directly proportional to the size of the simulated system (Table 1) as the degree of symmetry decreases upon the addition of new molecular objects. Figure 1 summarizes global-minimum molecular configurations of systems 3-5. We note that the major structural pattern is hydrogen bonding, H(MeOH)...O(DESO) whose length is 0.18 nm, whereas the oxygen-hydrogen-oxygen non-covalent angle varies from 171 to 176 degrees in different local-minimum configurations. A strong attraction between the sulfoxide molecule and the alcohol molecule is a cornerstone of their mutual miscibility.

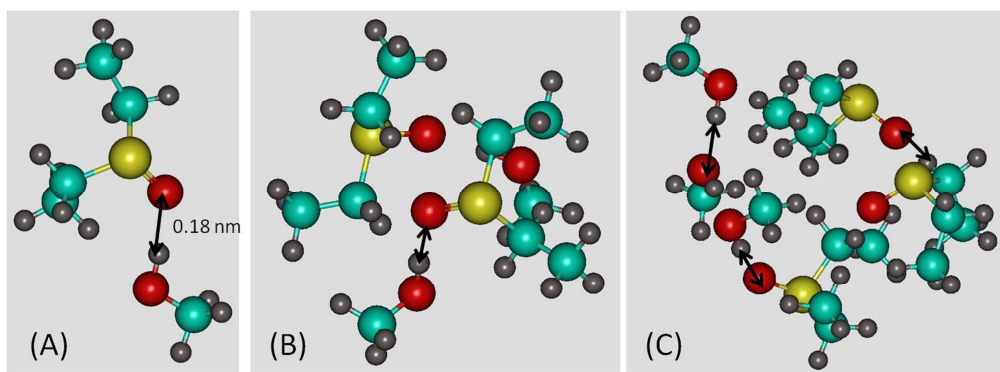

Figure 1. (A) Global-minimum molecular configuration of the DESO-MeOH pair (system 3). (B) Global-minimum molecular configuration of two DESO-MeOH pairs (system 4). (C) Global-minimum molecular configuration of three DESO-MeOH pairs (system 5). Non-covalent distances that determine the local structure of the mixture are marked with black arrows. The length of the O(DESO)...H(MeOH) hydrogen bond equals 0.18 nm and does not depend on the size of the cluster. Oxygen atoms are red, hydrogen atoms are grey, carbon atoms are cyan, sulfur atoms are yellow.

We found three substantially different minima states in system 3. All of them also have analogs in systems 4 and 5. Except for the hydrogen-bonded structural pattern depicted in Figure 1, an orientation of the polar moiety of MeOH to one of the alkyl chains of DESO occurs. This configuration is 32 kJ mol$^{-1}$ less favorable, as compared to the global-minimum molecular configuration. Next, one molecule (despite numerous efforts we could reproduce this behavior only for methanol) can leave the simulated cluster and go away as far as 0.9 nm. Further steepest-descent geometry optimization does not return this molecule to the condensed phase, therefore the resulting configuration must be considered an independent local minimum. The liquid-vapor interface is hereby formed. We calculated an energy cost of such a transition in system 3 to be +39 kJ mol$^{-1}$. The presented molecular configurations are not descriptive of the DESO-MeOH mixture, as confirmed by the energy of many k×T. We did not analyze them further.

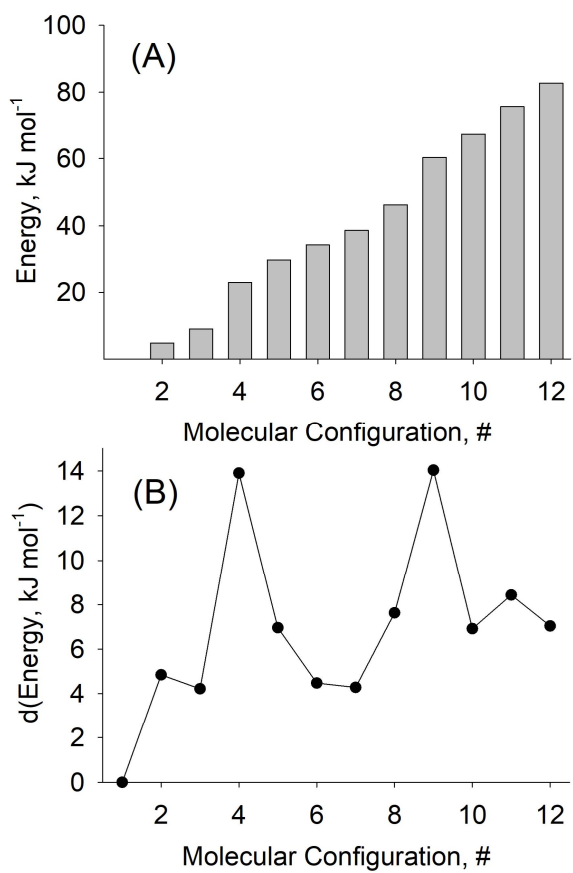

Figure 2. (A) Relative potential energies corresponding to local-minimum molecular configurations of system 6. (B) The potential energy difference between minima. All minima were arranged according to their increasing potential energies to simplify analysis. The global minimum was assigned zero energy.

Figures 2-3 depict distributions of local minima by energy and exemplify a few most representative structures. The energy difference between the global-minimum molecular configuration and the least thermodynamically stable local-minimum molecular configuration amounts to 83 kJ mol$^{-1}$. Therefore, significant conformational flexibility was detected in the DESO-MeOH system. Note that the stability of a certain molecular configuration depends on the shape of the potential energy surface, but does not directly depend on the local-minimum energy. The high-energy local minima patterns are highly unlikely to be found at low temperatures, but they must be expected in the system upon boiling and vaporization.

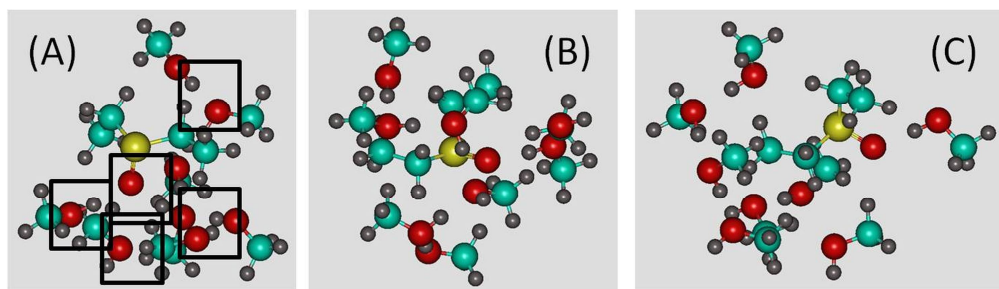

Figure 3. The most representative optimized configurations of system 6. (A) Global-minimum configuration. Hydrogen-bonded atoms are marked by black squares. (B) Arbitrarily chosen local-minimum configuration that is less thermodynamically stable than the global-minimum configuration by +30 kJ mol$^{-1}$. (C) The least thermodynamically stable local minimum with the relative potential energy of +83 kJ mol$^{-1}$. Oxygen atoms are red, hydrogen atoms are grey, carbon atoms are cyan, sulfur atoms are yellow.

All low-energy DESO-MeOH configurations (Figure 3) exhibit hydrogen bonding, H(MeOH)...O(DESO). Importantly, numerous hydrogen bonds are formed also between MeOH molecules, O(MeOH)...H(MeOH), with a length of 0.18 nm and an oxygen-hydrogen-oxygen non-covalent angle of more than 175 degrees. The hydrogen-bonded network created by MeOH and including the DESO molecule greatly stabilizes the system. For instance, no evaporation events were detected, whereas they sporadically occurred in other systems. We performed an additional investigation by exercising more aggressive and frequent kinetic energy injections but were unable to break the H-bonded network. This finding is essential to prove that dilute solutions of DESO in MeOH are thermodynamically very stable systems.[20] Hydrogen bonding is known to play an important role in the dimethyl sulfoxide – methanol mixtures which have been quite comprehensively investigated in numerous previous works.[5,7,23,25,43–45] In turn, the presence of the methylene groups in the structure of DESO, as opposed to the one of DMSO, adds conformational flexibility to the mixtures and, therefore, engenders a different set of physical-chemical properties.

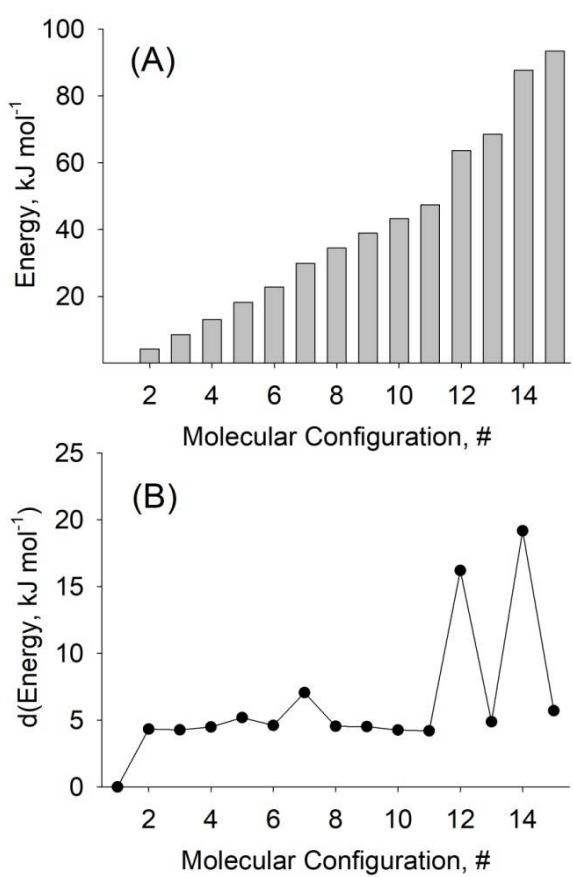

Figure 4. (A) Relative potential energies corresponding to local-minimum molecular configurations of system 7. (B) Potential energy difference between local-minimum molecular configurations. All minima were arranged according to their increasing energies. Each consequent minimum is less stable than the previous one. The global minimum was assigned zero energy.

The total number of local-minimum molecular configurations in system 7 (Figure 4) is smaller than in system 6 (Figure 2). This can be rationalized by a somewhat larger size of system 7 and higher conformational flexibility of the DESO molecule. The difference in energy between the global-minimum configuration and the least thermodynamically stable local minimum is also slightly larger, 93 kJ mol$^{-1}$ being in line with an observation of three more local minima. The most flexible one is system 5 (Table 1). Its versatility should be primarily linked to low symmetry.

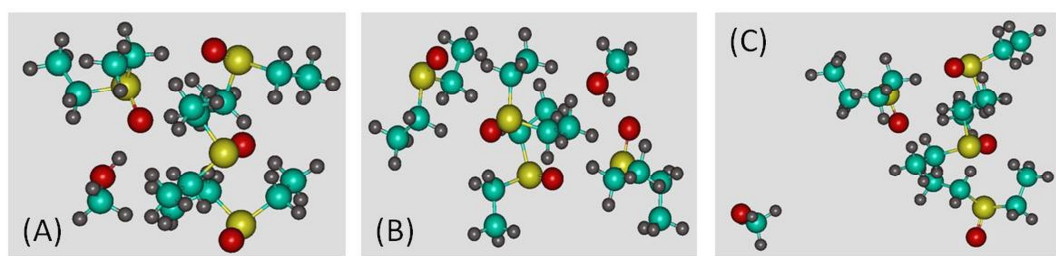

Figure 5. The most representative optimized molecular configurations of system 7. (A) Global-minimum configuration. (B) Very stable local minima configuration being 18 kJ mol$^{-1}$ less thermodynamically stable as compared to the global-minimum configuration. (C) The liquid-vapor local-minimum molecular configuration with the relative potential energy of +93 kJ mol$^{-1}$. Oxygen atoms are red, hydrogen atoms are grey, carbon atoms are cyan, sulfur atoms are yellow.

The MeOH molecule is located on the surface of the "1 MeOH + 4 DESO" system (Figure 5 A, B). Although it does create H-bond with the neighboring DESO molecule in most detected configurations, the location on the surface means that MeOH is not a solvate forming particle in its systems with DESO. The location of MeOH at the center of the cluster would mean the opposite. Figure 5C possesses the relative energy of +93 kJ mol$^{-1}$ and an unusual distance between the MeOH molecule and the closest atom of the cluster, ~1.0 nm. As a result of kinetic energy injection, virtual evaporation took place. The hydrogen bond with DESO was broken and the MeOH molecule drifted away. In other terms, it formed a gaseous phase. It is important to note that the absence of other MeOH molecules, that would maintain a hydrogen-bonded network, fostered this unwanted evaporation.

We computed RDF for the H(MeOH)...O(DESO) atom pair (Figure 6A) to confirm the intermolecular hydrogen bond existence based on the time-averaged space-averaged data. Furthermore, we computed RDF for the non-polar moieties of both molecules (Figure 6B) to prove that no spatial correlations exist in the hydrophobic regions of the mixture. The RDFs computed for the hydrogen-bonded atom pair exhibit a strong peak at 0.18 nm, whereas RDF for non-polar moieties is noisy. Interestingly, the correlation of the methanol's methyl group is stronger with the methylene groups of DESO compared to the methyl groups because the methylene group is less flexible thanks to its vicinity to the sulfoxide group.

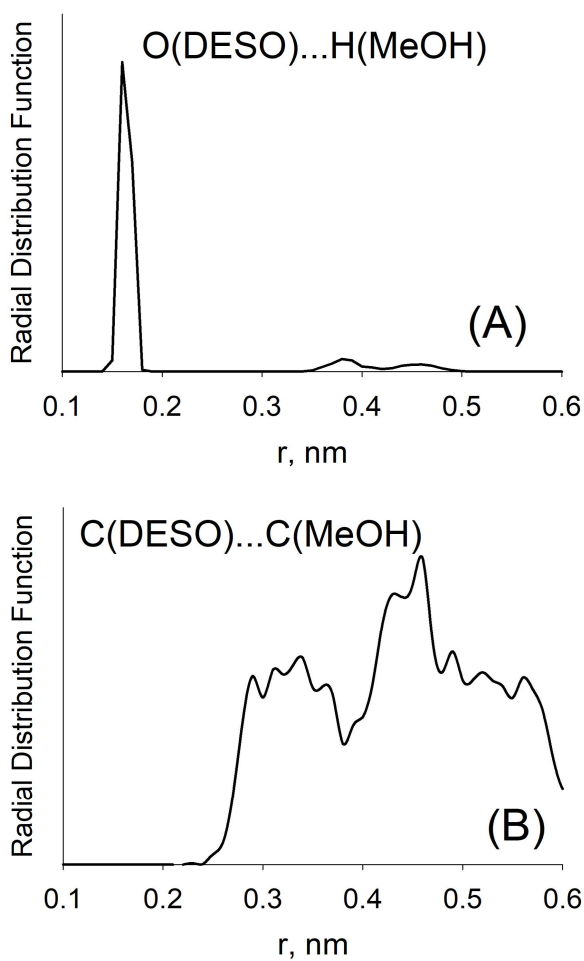

Figure 6. RDFs for the selected non-bonded atom pairs. (A) The oxygen atom of DESO and the methyl hydrogen atom of MeOH create a hydrogen bond that is 0.18 nm long. (B) Methyl group carbon atom of MeOH and methyl and methylene group carbon atom of DESO.

Figure 7 represents vibrational spectra of MeOH, DESO, and their equimolar mixture, DESO-MeOH. We identified the following major peaks. The sharp peak at 723 cm$^{-1}$ corresponds to wagging vibrations of O-H methanol. The peaks at 965 and 999 cm$^{-1}$ correspond to carbon-carbon and sulfur-oxygen symmetric stretching vibrations in the DESO molecule. At 1065 cm$^{-1}$, C-H of the methyl group of DESO exhibits rocking vibrations. The C-O bond of methanol exemplifies stretching and scissoring vibrations at 1067 and 1134 cm$^{-1}$, respectively. Scissoring vibrations are also seen for C-H of DESO, C-H of MeOH, and O-H of MeOH at 1446-1497 cm$^{-1}$. At 2974 and 3015 cm$^{-1}$ C-H of MeOH contributes symmetric and anti-symmetric stretching vibrations, respectively. At 3036-3048 cm$^{-1}$ and 3105 cm$^{-1}$, the symmetric stretching of C-H of DESO is seen. At 3095 and 3125 cm$^{-1}$, C-H anti-symmetric stretching of DESO was revealed. At 3541 cm$^{-1}$, the oxygen-hydrogen bond of the hydroxyl group of MeOH exhibits stretching vibrations.

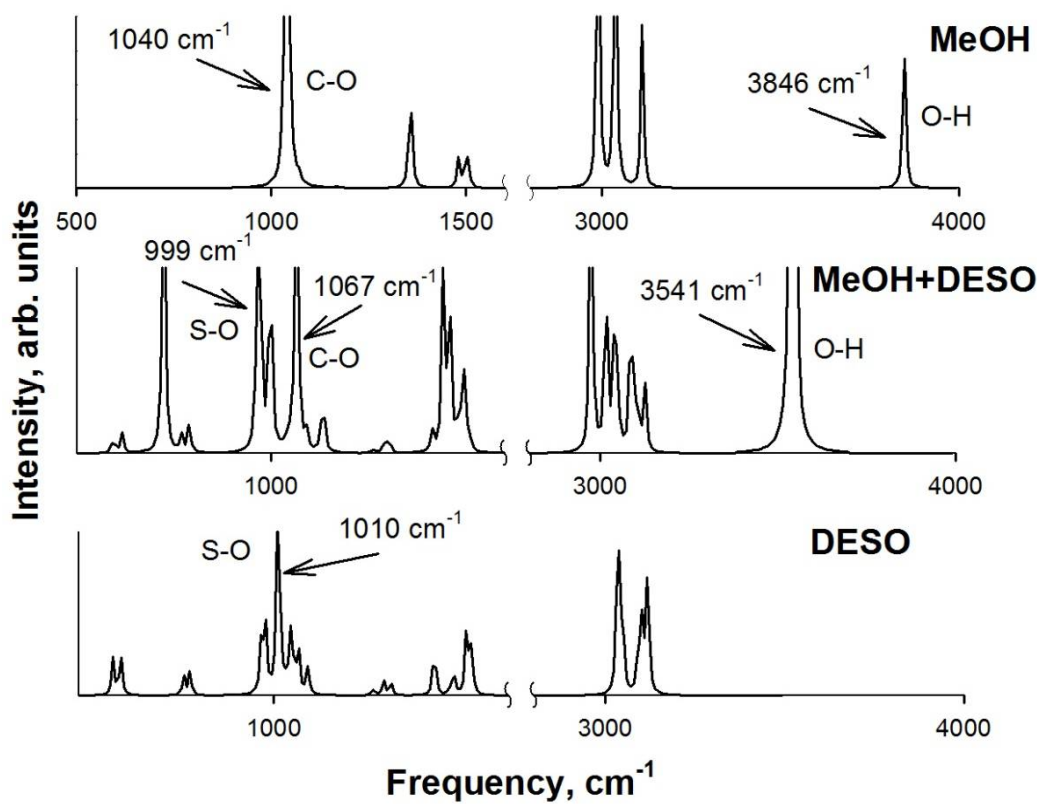

Figure 7. Mid- and far-infrared (IR) vibrational spectra of methanol (MeOH), diethyl sulfoxide (DESO), and the DESO-MeOH molecular pair. All IR spectra were computed for the global-minimum configurations at the B3LYP/6-311++G** level of theory.

We compared the computed infrared spectra with the experimentally obtained reference data available in the NIST WebBook.[46] In particular, the experimental carbon-oxygen band in MeOH is wide, 1012-1062 cm$^{-1}$. This peak is in perfect agreement with the computed frequency of 1040 cm$^{-1}$. In turn, the covalent oxygen-hydrogen vibration band is located at 3650-3750 cm$^{-1}$ in the experiment. This frequency is a few percent smaller than the computed peak at 3846 cm$^{-1}$. The sulfur-oxygen covalent vibration is delocalized between 1062-1125 cm$^{-1}$. It is only insignificantly redshifted in our calculations, 1010 cm$^{-1}$. In general, the computations carried out at the hybrid density functional level of theory with a significantly large set of basis functions, B3LYP/6-311++G**, provide reliable frequencies in the infrared range.

Significant displacements of certain peaks are observed in the spectrum of the mixture occurring thanks to a strong hydrogen bond, H (MeOH)...O (DESO), formation. The length of the hydrogen bond amounts to 0.18 nm. First, the peak corresponding to the S-O stretching vibration gets shifted towards a lower frequency. The peak at 1010 cm$^{-1}$ corresponds to pure DESO, whereas the peak at 999 cm$^{-1}$ corresponds to the mixture. The peak's blue shift thanks to mixing, therefore, equals 11 cm$^{-1}$. Second, the sharp peak corresponding to the C-O stretching vibration gets shifted from 1040 cm$^{-1}$ in pure MeOH to 1067 cm$^{-1}$ in the DESO-MeOH mixture. The displacement equals 27 cm$^{-1}$. Third, the peak corresponding to the O-H stretching vibration in pure MeOH gets shifted from 3846 cm$^{-1}$ to 3541 cm$^{-1}$ in the DESO-MeOH mixture. The blue shift is very significant, 305 cm$^{-1}$. Expectedly, all changes in the DESO-MeOH spectrum reflect strong non-covalent interactions between these two molecules due to a hydrogen bond formation. Thereby the analysis of the vibrational frequencies in the global-minimum DESO-MeOH molecular configuration confirms the conclusions that were derived upon the analysis of the local-minimum configurations and distribution functions.

**Conclusions**

Hereby we comprehensively analyzed molecular configurations that correspond to local and global minima on the potential energy surface of the systems comprising methanol and diethyl sulfoxide. We identified the global-minimum configurations in a range of DESO-MeOH mixtures, including the infinitely dilute solutions (one molecule of DESO and one molecule of MeOH, respectively). A strong hydrogen bond between the polarized hydroxyl hydrogen atom of the alcohol and the sulfoxide group oxygen atom of DESO was detected in all detected molecular configurations irrespective of their size and compositions. The length of the hydrogen bond, 0.18 nm, suggests that it makes a major contribution to the mutual solubility of DESO and MeOH. Furthermore, MeOH molecules retain a network of strong hydrogen bonds, with a length of 0.18 nm, thus forming a cage around the solvated DESO molecule. Therefore, a solvation mechanism of DESO in MeOH resembles the one outlined for the case of DESO-water mixtures.[15,16]

Radial distribution functions and a collection of local-minimum configurations along with the global-minimum structure work together in describing the condensed-state peculiarities. Whereas the RDF reflects the most common molecular configurations and atom-atom spatial correlations at some temperature and pressure conditions, low-energy local minima provide information about all possible mutual orientations in the system. Arranging local minima according to their potential energies allows understanding which configurations are most probable in the mixture.

By determining the most likely structural pattern in the DESO-MeOH mixtures, we unambiguously rationalized the dependence of volumetric properties vs. molar composition reported by Markarian and coworkers.[20] By using the infrared spectrum of the DESO-MeOH mixture we confirmed the formation of the strong O...H intermolecular hydrogen bond and characterized it.

**Acknowledgments**

**Conflict of interest**

The authors hereby declare no existing financial interests concerning these research studies.

**Authors Contributions**

V.V.C. analyzed available literature, formulated the research schedule, conducted simulations, summarized results, provided their discussion, and wrote the manuscript. N.A.A. computed infrared spectra and compared them to the experiment.

**Authors for correspondence**

All correspondence regarding the content of this paper shall be directed through electronic mail to the authors mentioned below:

V.V.C.: vvchaban@gmail.com.